\documentstyle[psfig]{mn}
\def\P3M{P$^3$M}
\def\and{, }
\font\mgn=cmti7
\def\eqnote#1{\marginpar{\mgn #1}}
\def\eqnote#1{{}}
\newfam\cyrfam
\font\tencyr=wncyr10 at 12pt
\font\sevencyr=wncyr7 at 8.5pt \font\fivecyr=wncyr5 at 6pt

\textfont\cyrfam=\tencyr \scriptfont\cyrfam=\sevencyr
            \scriptscriptfont\cyrfam=\fivecyr

\def\fracnum#1#2{\raise 2.1pt\hbox{$\scriptstyle #1$}\kern
-1.2pt/\kern -1.2pt \lower 2.1pt\hbox{$\scriptstyle#2$}\,}

\begin{document}

\title{Significant effects of weak gravitational lensing on determinations of the cosmology from Type Ia Supernov\ae}
 
\author[Andrew J. Barber] {Andrew J. Barber$^1$\thanks{Email: 
abarber@star.cpes.susx.ac.uk}\\ {}$^1$Astronomy Centre, University of
Sussex, Falmer, Brighton, BN1 9QJ\\ }

\date{Accepted 1999 ---. Received 1999 ---; in original form 1999 ---}

\maketitle 

\begin{abstract}

Significant adjustments to the values of the cosmological parameters
estimated from high-redshift Type Ia Supernov\ae~data are reported,
almost an order of magnitude greater than previously found. They arise
from the effects of weak gravitational lensing on observations of
high-redshift sources. The lensing statistics used have been obtained
from computations of the three-dimensional shear in a range of
cosmological $N$-body simulations, from which it is estimated that
cosmologies with an underlying deceleration parameter $q_0 = -0.51
+0.03/-0.24$ may be interpreted as having $q_0 = -0.55$ (appropriate
to the currently popular cosmology with density parameter $\Omega_M =
0.3$ and vacuum energy density parameter $\Omega_{\Lambda} = 0.7$). In
addition, the standard deviation expected from weak lensing for the
peak magnitudes of Type Ia Supernov\ae~at redshifts of 1 is expected
to be approximately 0.078 magnitudes, and 0.185 magnitudes at redshift
2. This latter value is greater than the accepted intrinsic dispersion
of 0.17 magnitudes. Consequently, the effects of weak lensing in
observations of high-redshift sources must be taken properly into
account.

\end{abstract}


\section{INTRODUCTION}

\subsection{Background}

The weak gravitational lensing of light from distant sources by the
large scale structure in the universe results in the appearance of
shear and convergence in images. The application of `ray-tracing'
methods, and other programmes, to cosmological $N$-body simulations has
enabled weak lensing statistics to be recorded for imaginary sources
at various redshifts. These statistics in general show significant
differences between different cosmological models, and, of course,
differences for sources at different redshifts.

From weak lensing magnifications computed from `ray-tracing' in
$N$-body simulations, Wambsganss et al. (1997) were able to estimate
the likely magnification biases for sources at redshifts $z=0.5$ and $z=1$, and
suggested that these biases should be applied to the observed
magnitudes of high-redshift Type Ia Supernov\ae. Their work was
conducted in a cosmological simulation with density parameter
$\Omega_M = 0.4$ and vacuum energy density parameter $\Omega_{\Lambda} = 0.6$,
for which the deceleration parameter is $q_0 = -0.4$. They found for
this cosmology that if the computed median demagnification for $z = 1$
were applied to the observed Supernov\ae~at this redshift it would shift
their positions on the Hubble diagram indicative of a different
cosmology with $q_0 = -0.395$.

Riess et al. (1998) and Perlmutter et al. (1999) have independently
used data sets of Type Ia Supernov\ae~extending to a redshift of 0.97
to estimate values of the cosmological parameters from the Hubble
diagram. In discussion of their results in the light of the work by
Wambsganss et al. (1997), Riess et al. (1998) have stated that the
effects on values of the cosmological parameters should be
negligible. Perlmutter et al. (1999) have assumed that the effects of
magnification or demagnification average out, and that the most
overdense (or high magnification) lines of sight should be rare for
their set of 42 high-redshift Supernov\ae.

However, Barber et al. (2000) have now applied a new algorithm
developed by Couchman, Barber and Thomas (1999) for the
three-dimensional shear to obtain weak lensing statistics in
simulations with cosmological parameters similar to those estimated
from the high-redshift Supernov\ae~data. Their results show that the
median demagnification values are considerably less and the
dispersions in the probability distributions for the magnification are
greater than found by Wambsganss et al. (1997) for redshifts of 1 and
0.5 (although from different cosmologies). This suggests that the work
of Wambsganss et al. (1997) should be reviewed in the light of the new
weak lensing data now available from a suitable cosmology.

In this paper I apply the new data from Barber et al. (2000) in an
attempt to re-estimate the effects of weak lensing on the cosmological
parameters, and find the effects to be far from negligible.

\subsection{Outline of paper}

A brief outline of this paper is as follows.

In {\bf Subsection 1.3} work by others on adjustments to the
cosmological parameters resulting from weak gravitational lensing is
summarised.

{\bf Section 2} summarises the method followed by Barber et al. (2000)
for obtaining the weak lensing statistics. {\bf Subsection 2.1}
describes the cosmological $N$-body simulations; {\bf Subsection 2.2}
summarises the main features of the algorithm used for the
determination of the shear in three-dimensions; {\bf Subsection 2.3}
outlines the method for integrating the shear values throughout the
linked simulation volumes, and describes how these values are then
used to determine the probability distributions for the magnifications
of sources at high-redshift.

{\bf Section 3} summarises the relevant weak lensing statistics which
are used in the present work.

In {\bf Section 4}, I describe the form of the Hubble diagram and its
dependence on the cosmological parameters, and also summarise the
important work by Riess et al. (1998) and Perlmutter et al. (1999) in
estimating the cosmological parameters from high-redshift Type Ia
Supernov\ae~data.

In {\bf Section 5}, I discuss how the dispersions in the magnification
distributions for different source redshifts may affect the observed
distance moduli. For the specific case of the $\Omega_M = 0.3$,
$\Omega_{\Lambda} = 0.7$ cosmology, I then consider by how much the
Supernov\ae~data are displaced on the Hubble diagram, and how this
translates into effective changes to the perceived values of the
cosmological parameters.

{\bf Section 6} summarises the results obtained for the deceleration
parameter, and discusses the findings in the light of work by others.

\subsection{Other work}

Wambsganss et al. (1997) have used the same ray-tracing method as
described by Wambsganss, Cen and Ostriker (1998) for cosmological
simulations with $\Omega_M = 0.4$, $\Omega_{\Lambda} = 0.6$, and
normalisation $\sigma_8 = 0.79$. The magnification values above and
below which 97\mbox{$\frac{1}{2}$}\% of all of their lines of sight
fall were $\mu_{\mathrm{low}}=0.951$ and $\mu_{\mathrm{high}}=1.101$
for source redshifts of $z = 1$, and $\mu_{\mathrm{low}}=0.978$ and
$\mu_{\mathrm{high}}=1.034$ for $z=0.5$. The median values for the
magnification were 0.983 at $z = 1$, and 0.993 at $z = 0.5$. The
authors claim that these values would give rise to observed values of
$q_0 = -0.395 +0.125/-0.095$ for the $z = 1$ data, and $q_0 = -0.398 +
0.048/-0.077$ for the $z = 0.5$ data, rather than the assumed value of
$-0.4$. (The quoted errors arise from the magnification ranges
described above in the asymmetrical distributions.) They also report
that the lensing-induced dispersions in their critical density
cosmology are three times larger, but this cosmology uses a
normalisation of $\sigma_8 = 1.05$ which overproduces the present-day
rich cluster abundances. For this cosmology, they report
magnifications up to 100, and correspondingly highly dispersed
distributions, very much larger than those of Barber et al. (2000) for
$z=3$. The high magnifications almost certainly derive from the low
value of their (fixed) softening scale.

Fluke, Webster and Mortlock (1999) and Fluke, Webster and Mortlock
(2000) have used a `ray-bundle' method in which a discrete bundle of
light rays is traced. The method allows a direct comparison between
the shape and size of the bundle at the observer and at the source
plane, so that the magnification, ellipticity and rotation can be
determined straightforwardly. The cosmological models investigated
were similar to those of Barber et al. (2000). However, the particles
were considered as point masses, so that very high magnification
values could be achieved in principle; however, to alleviate this
situation, the authors do not include bundles which pass within
$\sqrt{2}$ of the Einstein radius of any particle.  Because of their
use of point particles, they use the empty beam approximation which
gives rise to magnification probability distributions with minimum
magnifications of 1 (and therefore mean and median values greater
than, or equal to, one), high magnification tails, and broad
dispersions in magnification.  For $z = 1$, the distribution in the
magnifications for the SCDM cosmology ($\Omega_M = 1,$
$\Omega_{\Lambda} = 0$) is clearly broader than that for the low
density (LCDM) cosmology ($\Omega_M = 0.3$, $\Omega_{\Lambda} = 0.7$),
which in turn is broader than the open (OCDM) cosmology ($\Omega_M =
0.3,$ $\Omega_{\Lambda} = 0$). This is also true of Barber et al.'s
(2000) data, although Barber et al. (2000) find a completely different
order for the cosmologies at higher source redshifts. Because of the
high magnification tails in Fluke, Webster and Mortlock's (2000)
distributions, they discard those high magnification lines of sight
which occur with low probability before defining the
$\mu_{\mathrm{low}}$ and $\mu_{\mathrm{high}}$ values. 

These magnification probability distributions are then used by Fluke
and Webster (2000) to examine the effects of the weak lensing
dispersion on measurements of $q_0$ in both the empty beam (more
appropriate for point particles) and full beam limits.  They consider
whether the resulting shifts in the Hubble diagram can be fitted with
various cosmological models.  With input data from the $z=1$
magnification distribution for the flat LCDM simulations, they find
that $q_0 = -0.53 +0.96/-0.02$ from the empty beam distribution.  To
convert their empty beam magnification values to filled beam values,
Fluke and Webster (2000) use a simple scaling relationship. However,
since the filled beam approach is inappropriate to their method, the
resulting values should be treated with caution. They find the revised
value $q_0 = -0.61~+0.89/-0.02$; i.e., a much larger departure than
for the empty beam approximation, but a similar dispersion. Again the
dispersion can be partly explained in terms of their use of point
particles, although they do attempt to modify their distributions to
account for the high magnification tails.

It should be mentioned that Wambsganss et al. (1997) have assumed that
the observed Supernov\ae~have magnifications of unity, and then have
assessed departures from the best fit cosmology, rather than assuming
that the best fit cosmology comes from the most likely demagnification
at the peak of the magnification distributions, which is the approach
I take in this paper. Although Fluke and Webster (2000) use the median
magnification values, the values are always close to unity and greater
than, or equal to, unity for the empty beam case.

\section{METHOD}

\subsection{The Hydra {\em N}-body simulations}

The cosmological $N$-body simulations used in the study by Barber et
al. (2000) were provided by the Hydra Consortium
\footnote{http://hydra.mcmaster.ca/hydra/index.html} and produced
using the `Hydra' $N$-body hydrodynamics code (Couchman, Thomas and
Pearce, 1995). Results are reported on simulations from four different
cosmologies, referred to as the SCDM, TCDM, OCDM and LCDM
cosmologies. Each of the simulations uses a Cold Dark Matter-like
spectrum, and the parameters used in the generation and specification
of these simulations are listed in Table~\ref{cosmo}. $\Omega_M$ and
$\Omega_{\Lambda}$ are the present-day values of the density parameter
and the vacuum energy density parameter respectively. The power
spectrum shape parameter, $\Gamma$, is set to 0.5 in the SCDM
cosmology, but the empirical determination (Peacock and Dodds, 1994)
of 0.25 for cluster scales has been used in the other cosmologies. In
each case, the normalisation, $\sigma_8$, on scales of $8h^{-1}$Mpc
has been set to reproduce the number density of clusters (Viana and
Liddle, 1996); $h$ is the Hubble parameter expressed in units of
100km~s$^{-1}$Mpc$^{-1}$.
\begin{table}
\begin{center}
\begin{tabular}{|c|c|c|c|c|c|c|}
\hline
Cosmology & $\Omega_M$ & $\Omega_{\Lambda}$ & $\Gamma$ & $\sigma_8$ & No. of    & Box side \\
          &            &                    &          &            & particles & ($h^{-1}$Mpc) \\   
\hline 
SCDM & 1.0 & 0.0 & 0.50 & 0.64 & $128^3$ & 100 \\ 
\hline 
TCDM & 1.0 & 0.0 & 0.25 & 0.64 & $128^3$ & 100 \\ 
\hline 
OCDM & 0.3 & 0.0 & 0.25 & 1.06 & $86^3$ & 100 \\ 
\hline 
LCDM & 0.3 & 0.7 & 0.25 & 1.22 & $86^3$ & 100 \\
\hline
\end{tabular}
\end{center}
\caption{Parameters used in the generation of the four different cosmological simulations.}
\label{cosmo}
\end{table}

In the SCDM and TCDM cosmologies, the number of particles was $128^3$,
leading to individual dark matter particle masses of $1.29 \times
10^{11}h^{-1}$ solar masses. In the low density universes, the number
of particles was 0.3 times the number in the critical density
universes, leading to the same individual particle masses. The
simulation output times were chosen so that consecutive simulation
volumes could be snugly abutted; the side-dimensions were $100h^{-1}$Mpc in
every case. 

Since each time-output in a given simulation run is generated using
the same initial conditions, particular structures (although evolving)
occur at the same locations in all the boxes, and are therefore
repeated with the periodicity of the box. To avoid these correlations,
each simulation box was arbitrarily translated, rotated (by
multiples of $90^{\circ}$) and reflected about each coordinate axis,
before linking together to form the continuous depiction of the
universe back to the source redshift.

\subsection{The three-dimensional shear algorithm}

Couchman et al. (1999) describe in detail the algorithm for the
three-dimensional shear which has been applied by Barber et al. (2000)
to the simulations for the different cosmologies. The main features of
the code are as follows.

First, the algorithm uses a Fast Fourier Transform method in the
Particle-Mesh part of the code, in which the density distribution is
smoothed. The procedure makes use of the periodicity of the
fundamental volume, so that the effects of matter effectively
stretching to infinity are included in the computations. Couchman et
al. (1999) pointed particularly to the importance of including the
effects of matter beyond a single period orthogonal to the direction
of the line of sight.

Second, in order to evaluate the shear components correctly, the
algorithm works with the `peculiar gravitational potential,' which
describes departures in the potential from homogeneity, and which
results in net zero mass for the universe.

Third, a key feature of the algorithm is the variable particle
softening. The feature enables particles in low density regions to
have extended softening, so that nearby evaluation positions for the
shear register a density rather than a complete absence of matter. By
contrast, particles in highly clustered regions are assigned low
softening values, and a selected minimum value is introduced which
limits the possibility of strong lensing behaviour.  The minimum value
of the softening in the work by Barber et al. (2000) was selected to
be 10$^{-3}$ in box units, equivalent to 0.1$h^{-1}$Mpc at $z=0$.  The
variable softening feature enables a much more realistic depiction of
the large-scale structure within a simulation to be made.

The output from the algorithm is the six independent components of the
three-dimensional shear evaluated at a large number of positions
within each simulation cube. 

\subsection{Method}

A regular rectangular grid of $100 \times 100$ directions through each
box was established. (Since there were only small deflections, and the
point of interest was the statistics of output values, each light ray
was considered to follow one of the lines defined by these directions
through the boxes.) Each `ray' was then connected with the
corresponding line of sight through subsequent boxes in order to
obtain the required statistics of weak lensing. In the application of
the code, the shear was evaluated at 1000 positions along each of the
lines of sight, forming a regular grid of evaluation positions in each
box.

The evaluation of the three-dimensional shear within the volume of the
boxes, used the appropriate angular diameter distances at every
evaluation position, representing a distinct advantage over
two-dimensional (planar) methods. Furthermore, the variable softening
facility within the algorithm lead naturally to the assumption that
the universe may be described in terms of the filled beam approach,
and this was the approach adopted by Barber et al. (2000). This is
different from the assumptions of many other workers who frequently
use point particles, or a limited form of fixed softening, and use the
empty beam approximation. The two approaches give rise to quite
different expectations and results, the most obvious being the
following. First, strong lensing can occur with particles of small
radii, leading to high magnification tails in the probability
distributions. Second, magnification distributions in the empty beam
approximation all have minimum magnifications of $\mu_{\mathrm {min}}
= 1$, whilst in the filled beam, $\mu_{\mathrm {min}} \leq 1$; this
may alter the dispersions and the median values in the two
distributions, essential to an understanding of the magnitudes
expected for high-redshift Type Ia Supernov\ae. Finally, the mean
values for the magnifications can be calculated from the respective
angular diameter distances in the different cosmologies, for the empty
beam approximation; however, the mean values in the filled beam
approximation are always 1.

The six independent second derivatives of the peculiar gravitational
potential were calculated by the code at each of the selected
evaluation positions throughout each box.  These were then integrated
in small steps along each line, forming, essentially, a large number
of deflection sites through each simulation box.  The integrated
values formed the input data to establish the elements of the
two-dimensional Jacobian matrix on each of the lines of sight for each
of the deflection sites. The Jacobian matrix elements were then used
together with the multiple lens-plane theory (summarised by Schneider,
Ehlers and Falco, 1992) and the appropriate angular diameter distances
to obtain values for the magnification, source ellipticity, shear and
convergence from each of the 10,000 lines of sight throughout all the
simulation boxes linked back to the required redshift.

In the procedure to obtain the two-dimensional second derivatives of
the effective lensing potentials, it was assumed that (a) the angular
diameter distances varied linearly between the evaluation step points
(although they were evaluated exactly at each step point), (b) the
angular diameter distances varied continuously through the depth of
each simulation box, as they would in the real universe, (c) the
shearing of light was weak, so that `rays' were considered to follow
the straight lines of sight defined by the grid of evaluation
positions within each simulation box, and (d) the accumulating
component values within the developing Jacobian matrices were computed
using the assumption that the shear was weak.

\section{WEAK LENSING IN DIFFERENT COSMOLOGIES}

Barber et al. (2000) have reported significant ranges of
magnifications in all the cosmologies for nominal source redshifts of
0.5, 1.0, 2.0, 3.0 and 4.0. From the magnification distributions they
have computed the median values, $\mu_{\mathrm{peak}}$, the values,
$\mu_{\mathrm{low}}$ and $\mu_{\mathrm{high}}$, above and below which
97\mbox{$\frac{1}{2}$}\% of all lines of sight fall, and also the rms
deviations from unity for the magnifications. All the values mentioned
are displayed in Table~\ref{mu4C}, which is reproduced from Barber et
al. (2000).
\begin{table}
\begin{center}
\begin{tabular}{|c|c|c|c|c|}
\hline
$z$ & $\mu_{\mathrm{low}}$ & $\mu_{\mathrm{peak}}$ &
rms deviation & $\mu_{\mathrm{high}}$ \\ \hline
\underline{SCDM} & & & & \\ 
3.9 & 0.835 & 0.933 & 0.115 & 1.420 \\
3.0 & 0.852 & 0.949 & 0.101 & 1.367 \\
1.9 & 0.885 & 0.947 & 0.079 & 1.277 \\
1.0 & 0.930 & 0.973 & 0.049 & 1.181 \\
0.5 & 0.969 & 0.985 & 0.023 & 1.089 \\ \hline
\underline{TCDM} & & & & \\ 
3.9 & 0.861 & 0.959 & 0.091 & 1.315 \\
3.0 & 0.877 & 0.951 & 0.081 & 1.286 \\
1.9 & 0.904 & 0.966 & 0.064 & 1.228 \\
1.0 & 0.941 & 0.972 & 0.039 & 1.144 \\
0.5 & 0.974 & 0.986 & 0.019 & 1.067 \\ \hline
\underline{OCDM} & & & & \\ 
4.0 & 0.858 & 0.919 & 0.115 & 1.469 \\
2.9 & 0.884 & 0.939 & 0.115 & 1.469 \\
2.0 & 0.915 & 0.942 & 0.069 & 1.283 \\
1.0 & 0.960 & 0.972 & 0.033 & 1.147 \\
0.5 & 0.985 & 0.989 & 0.013 & 1.062 \\ \hline
\underline{LCDM} & & & & \\ 
3.6 & 0.789 & 0.885 & 0.191 & 1.850 \\
2.0 & 0.870 & 0.934 & 0.108 & 1.453 \\
1.0 & 0.944 & 0.966 & 0.045 & 1.191 \\
0.5 & 0.981 & 0.987 & 0.016 & 1.070 \\ 
\hline
\end{tabular}
\end{center}
\caption{Various magnification statistics for the different cosmologies, as described in the text.}
\label{mu4C}
\end{table}

\begin{figure}
$$\vbox{
\psfig{figure=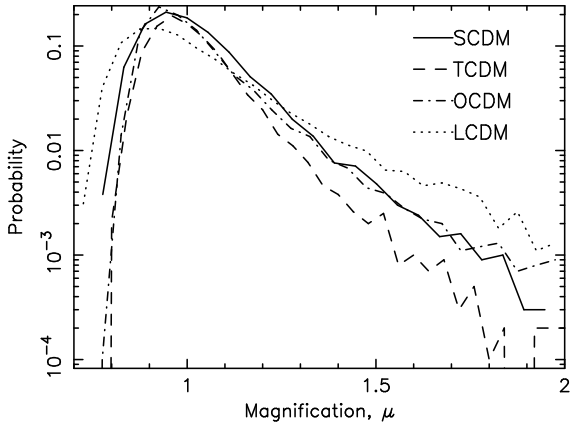,width=8.7truecm,angle=270}
}$$
\caption{The magnification probability distributions for all the cosmologies, assuming $z = 4$.} 
\label{magdistz4C}
\end{figure}

\begin{figure}
$$\vbox{
\psfig{figure=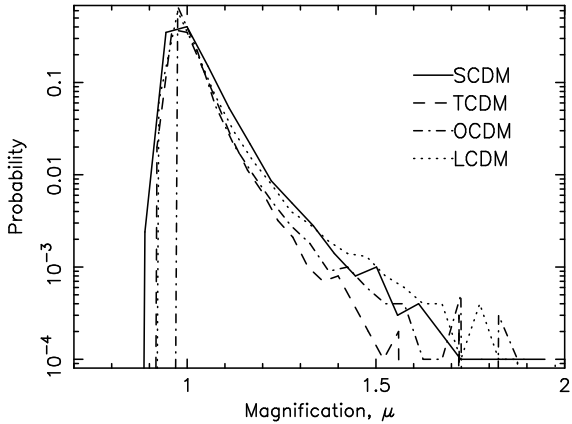,width=8.7truecm,angle=270}
}$$
\caption{The magnification probability distributions for all the cosmologies, assuming $z = 1$.} 
\label{magdistz1C}
\end{figure}

It is interesting to compare the distributions in the different
cosmologies. Figure~\ref{magdistz4C} and~\ref{magdistz1C} (also
reproduced from Barber et al., 2000) show the magnification
distributions for all the cosmologies for source redshifts of 4 and 1,
respectively. The distributions are all broader in the SCDM cosmology
when compared with the TCDM cosmology, due to its more clumpy
character. For source redshifts of 4 the OCDM and SCDM cosmologies
have very similar distributions even though the angular diameter
distance multiplying factors are larger in the OCDM cosmology. The
most significant feature for our purposes here is that for high source
redshifts the magnification distributions are broadest in the LCDM
cosmology (and the maximum values of the magnification are greatest
here), but for lower source redshifts the width of the distribution
falls below that for the SCDM and OCDM cosmologies. In the present
work, I shall be mostly concerned with the magnification values for
sources at $z = 1.0$ in the LCDM cosmology.

\section{\bf DETERMINATIONS OF THE COSMOLOGY FROM TYPE Ia SUPERNOV\AE}

The significant ranges in the magnifications (dependent on the
cosmology) which might apply to distant sources have been summarised
in the previous section. In the absence of magnification (or
demagnification) from the large-scale structure, it would be possible
to determine the cosmological parameters, $\Omega_M$ and $\Omega_{\Lambda}$,
from the departures from linearity in the Hubble diagram, provided
`standard candle' sources together with good calibration were
available for measurement at high redshift. This is precisely the
route taken by a number of authors, most importantly Riess et
al. (1998) and Perlmutter et al. (1999), both of whom have used
high-redshift Type Ia Supernov\ae~data of redshifts up to 0.97.

Carroll, Press and Turner (1992) give the general expression for the
distance measure to redshift, $z$:
\begin{eqnarray}
\lefteqn{\frac{H_0D_L}{c(1+z)}=\frac{1}{\mid\Omega_k\mid^{1/2}}\mathrm{sinn}\{\mid\Omega_k\mid^{1/2}\times} \nonumber\\
& & \int_{0}^{z}[(1+z')^2(1+\Omega_Mz')-z'(2+z')\Omega_{\Lambda}]^{-1/2}dz'\},
\label{Carroll}
\end{eqnarray}
where sinn means the hyperbolic sine, sinh, if $\Omega_k > 0$, or sine
if $\Omega_k < 0;$ $H_0$ is the Hubble parameter, $D_L$ is the luminosity distance, related to the
angular diameter distance through $D_L=(1+z)^2D$, and $c$ is the velocity of light; the curvature
density parameter, $\Omega_k$, is defined in terms of the matter
density parameter, $\Omega_M$, and the vacuum energy density
parameter, $\Omega_{\Lambda}$:
\begin{equation}
1=\Omega_M+\Omega_{\Lambda}+\Omega_k.
\end{equation}
However, if $\Omega_k=0$ (which I have assumed for this work), then
\begin{eqnarray}
\lefteqn{\frac{H_0D_L}{c(1+z)} = \int_{0}^{z}[(1+z')^2(1+\Omega_Mz')-z'(2+z')\Omega_{\Lambda}]^{-1/2}dz'} \nonumber\\
&\equiv& I,
\label{Carroll2}
\end{eqnarray}
and this distance measure, $I$, can be evaluated straightforwardly by
numerical integration for any cosmology and any redshift. At low
redshifts the expression reduces to the well-known Hubble
law, but departures from the linear Hubble relation become evident as
$z$ is increased, and then different cosmologies may be
differentiated.

Observationally, the distance modulus, $m-M$, (where $m$ is the
apparent magnitude, and $M$ the `corrected' absolute magnitude) is
determined, and this is related to the distance measure through
\begin{equation}
m-M=25+5{\mathrm{log}}_{10}[3000(1+z)I]-5{\mathrm{log}}_{10}h;
\label{modulus}
\end{equation}
see, e.g., Peebles (1993).

Consequently, the value of the distance modulus is fixed for any given
redshift, cosmology and Hubble parameter, but if the source is
magnified (or demagnified) by weak gravitational lensing effects, then
it will appear to be closer (or further away), according to
\begin{equation}
\Delta m = -2.5\mathrm{log}_{10}\mu,
\label{Deltam}
\end{equation}
where $\mu$ is the magnification. Alternatively, the observational
evidence could be interpreted as pointing to different values of the
cosmological parameters, $\Omega_M$ and $\Omega_{\Lambda}$, since the
curve on the Hubble diagram will be slightly displaced.

I now briefly summarise the work of Riess et al. (1998) and
Perlmutter et al. (1999), who have been able to estimate the
cosmological parameters from data-sets of high-redshift Type Ia
Supernov\ae.

Riess et al. (1998) have analysed the spectral and photometric
observational data of a total of 50 Type Ia Supernov\ae~with redshifts
$0.01 \leq z \leq 0.97$. They have made use of two different
correction methods for the peak magnitudes, namely the Multicolour
Light Curve Shape method (MLCS) and a template method. The MLCS
method, due to Riess, Press and Kirshner (1996) is an extension of
Riess, Press and Kirshner's (1995) empirical method, described above,
and makes use of up to 4 colours in the Supernov\ae~photometry to
quantify the amount of reddening by interstellar extinction; in
addition, Riess et al. (1998) have improved the method by using a
larger `training set' of Supernov\ae~and including a quadratic term in
the algorithm describing the light curve shapes. The template method is
that of Hamuy et al. (1995) combined with Phillips' (1993) light curve
decline parameter, $\Delta m_{15}(B)$. The fitting of the data has
been done using a $\chi^2$ statistic for a wide range of the
parameters, $H_0$, $\Omega_M$ and $\Omega_{\Lambda}$, and including
both the dispersions in the host galaxy redshifts (in terms of the
distance moduli), and the intrinsic dispersions (after correction) in
the Supernov\ae~peak magnitudes. With the constraint $\Omega_M +
\Omega_{\Lambda} = 1$, Riess et al. (1998) find that $\Omega_{\Lambda}
> 0$ at the $7\sigma$ and $9\sigma$ confidence levels for the two
correction methods, and formally quote, with this constraint,
\begin{equation}
\Omega_M = 0.28 \pm 0.10,~~~\Omega_{\Lambda} = 0.72 \pm 0.10
\label{cosmo1}
\end{equation}
using the MLCS method, and
\begin{equation}
\Omega_M = 0.16 \pm 0.09,~~~\Omega_{\Lambda} = 0.84 \pm 0.09
\label{cosmo2}
\end{equation}
using the template method.

Perlmutter et al. (1999) have based their analysis on 42 high-redshift
Type Ia Supernov\ae~having redshifts $0.18 \leq z \leq 0.83$, and
combined their results with 18 Type Ia Supernov\ae~from the
Cal\'{a}n/Tololo Supernova Survey having redshifts less than
0.1. Their method for determining the peak magnitudes relies on an
empirical relationship between the light-curve width and the
luminosity for the various template light-curves used. First, the time
axis of the template is dilated by the factor $(1+z)$ to allow for the
cosmological lengthening of the timescale. Then the observed
light-curve is fitted to the template by introducing a `stretch
factor,' $s$, on the time axis. Finally, the peak magnitude is
corrected to
\begin{equation}
m_{\mathrm{corr}} = m + \Delta_{\mathrm{corr}}(s),
\label{mcorr}
\end{equation}
where the correction term, $\Delta_{\mathrm{corr}}(s)$ is a simple
monotonic function of $s$. The correction terms (different for
different wavebands) were determined by Perlmutter et al. (1997) by
relationship to the decline $\Delta m_{15}$, and the assessment of the
luminosity by Hamuy et al.'s (1995) method. They claim that the residual
dispersion in the peak magnitudes for all the Supernov\ae, after applying the
correction based on the light-curve width-luminosity relationship, is
within about 0.17 magnitudes. They state, however, that it is not
clear whether the dispersion is best modeled as a normal distribution
in terms of the flux, or a log-normal distribution (Gaussian in
magnitude terms). This point is worth bearing in mind if Monte Carlo
simulations are performed to sample both the intrinsic dispersion and
the magnification distribution arising from weak lensing. The
preferred fitting of the data by Perlmutter et al. (1999) was
performed by minimising $\chi^2$ using the magnitude residuals in the
Hubble diagram, from which they find:
\begin{equation}
0.8\Omega_M - 0.6 \Omega_{\Lambda} \simeq -0.2 \pm 0.1.
\label{Perl1}
\end{equation}
With the constraint, $\Omega_M + \Omega_{\Lambda} = 1$, they then
find:
\begin{equation}
\Omega_M = 0.28^{+0.09}_{-0.08}(1\sigma\mathrm{statistical})^{+0.05}_{-0.04}(\mathrm{systematic}).
\label{Perl2}
\end{equation}

It is worth noting, however, that the above results are far from
conclusive. Goodwin et al. (1999) for example, have pointed to the
consistency of the data with a zero cosmological constant, provided
the Hubble parameter has a higher local value compared with the global
value; Riess at al. (1999) have pointed to the possible evolution of
Type Ia Supernov\ae~light-curve shapes arising from differences in the
rise-times, which may affect the estimation of the peak magnitudes,
although this has been questioned by Aldering, Knop and Nugent (2000).

\section{THE EFFECTS OF WEAK LENSING ON DETERMINATIONS OF THE COSMOLOGICAL PARAMETERS}

Both groups of workers, i.e., Riess et al. (1998) and Perlmutter et
al.  (1999), point to cosmologies which are close to the $\Omega_M =
0.3,$ $\Omega_{\Lambda} = 0.7$ cosmological simulation Barber et
al. (2000) analysed in terms of weak lensing. I shall therefore
discuss the dispersions in magnification and the impact on
determinations of the deceleration parameter, $q_0$, with regard to
this assumed cosmology.

\begin{figure}
$$\vbox{ \psfig{figure=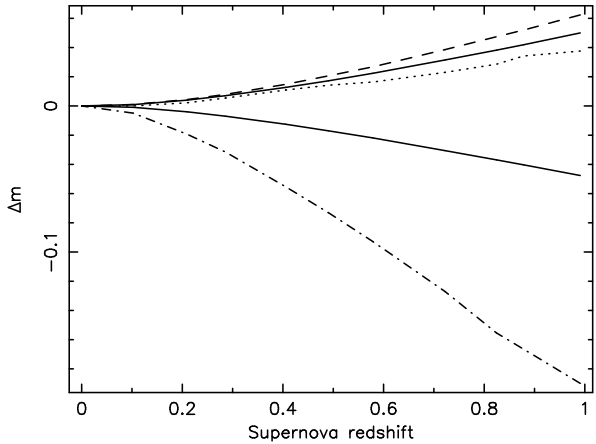,width=8.7truecm,angle=270} }$$
\caption{Differences in the distance modulus for appropriate values 
of the magnifications, as computed from the distribution curves for
Type Ia Supernov\ae~sources at redshifts $0 \leq z \leq 1$ in the LCDM
cosmology. The solid curves show the dispersion in distance modulus
arising from the rms fluctuations in magnification; the dotted line
shoes the departure in distance modulus arising from
$\mu_{\mathrm{peak}}$; the dashed line arises from
$\mu_{\mathrm{low}}$; and the dotted and dashed line arises from
$\mu_{\mathrm{high}}$.}
\label{mod.qdp}
\end{figure}

By considering Type Ia Supernov\ae~sources at ten evenly spaced
redshifts between $z = 0$ and $z = 1$, the resulting distributions of
the magnifications show the expected increasing dispersions, which
could be interpreted as observational dispersions in the peak
magnitudes. In Table~\ref{mus}, I give the values computed for
$\mu_{\mathrm{low}}$, $\mu_{\mathrm{peak}}$, the rms deviation in
magnification and $\mu_{\mathrm{high}}$, and, in Table~\ref{Deltams},
what these values mean in terms of differences in the distance modulus
compared with unmagnified sources. The final column in
Table~\ref{Deltams} gives the total difference in distance modulus
between the values derived from $\mu_{\mathrm{low}}$ and
$\mu_{\mathrm{high}}$. The values for the changes in the distance
modulus for appropriate values of the (de)magnification are shown in
Figure~\ref{mod.qdp}. The asymmetry in the magnification distributions
is clearly shown in terms of the distance modulus changes.

\begin{table}
\begin{center}
\begin{tabular}{|c|c|c|c|c|}
\hline
$z$ & $\mu_{\mathrm{low}}$ & $\mu_{\mathrm{peak}}$ &
rms deviation & $\mu_{\mathrm{high}}$ \\ \hline
0.99 & 0.944 & 0.966 & 0.045 & 1.191 \\
0.88 & 0.953 & 0.969 & 0.038 & 1.166 \\
0.82 & 0.957 & 0.974 & 0.035 & 1.154 \\
0.72 & 0.965 & 0.979 & 0.029 & 1.124 \\
0.58 & 0.976 & 0.985 & 0.020 & 1.089 \\
0.49 & 0.981 & 0.987 & 0.016 & 1.070 \\
0.41 & 0.986 & 0.990 & 0.012 & 1.052 \\
0.29 & 0.993 & 0.995 & 0.006 & 1.029 \\
0.21 & 0.996 & 0.998 & 0.004 & 1.018 \\
0.10 & 0.999 & 1.000 & 0.001 & 1.005 \\
\hline
\end{tabular}
\end{center}
\caption{The various magnifications measured from the distributions for Type Ia Supernov\ae~at the redshifts quoted in column 1. All the values have been determined for the LCDM cosmology.}
\label{mus}
\end{table}
\begin{table}
\begin{center}
\begin{tabular}{|c|c|c|c|c|c|}
\hline
$z$ & $\Delta m_{\mathrm{low}}$ & $\Delta m_{\mathrm{peak}}$  & $\Delta m_{\mathrm{rms}}$ & $\Delta m_{\mathrm{high}}$ & Dispersion \\ 
      &                           &                             &                           &                            & in $m$ \\
\hline
0.99 & 0.062 & 0.038 & -0.048/+0.050 & -0.190 & 0.252 \\
0.88 & 0.052 & 0.034 & -0.040/+0.042 & -0.167 & 0.219 \\
0.82 & 0.047 & 0.029 & -0.037/+0.038 & -0.155 & 0.203 \\
0.72 & 0.039 & 0.023 & -0.031/+0.032 & -0.127 & 0.166 \\
0.58 & 0.027 & 0.016 & -0.022/+0.022 & -0.093 & 0.120 \\
0.49 & 0.021 & 0.014 & -0.017/+0.017 & -0.073 & 0.094 \\
0.41 & 0.015 & 0.011 & -0.013/+0.013 & -0.055 & 0.070 \\
0.29 & 0.008 & 0.005 & -0.007/+0.007 & -0.032 & 0.039 \\
0.21 & 0.004 & 0.002 & -0.004/+0.004 & -0.019 & 0.024 \\
0.10 & 0.001 & 0.000 & -0.001/+0.001 & -0.005 & 0.006 \\
\hline
\end{tabular}
\end{center}
\caption{Differences in the distance modulus, when compared with unmagnified sources, for the appropriate computed magnifications from the distributions for Type Ia Supernov\ae~sources at the redshifts given in column 1. The values in the final column represent the total dispersion in the distance modulus for sources with (de)magnifications of $\mu_{\mathrm{low}}$ and $\mu_{\mathrm{high}}$. All the values have been determined for the LCDM cosmology.}
\label{Deltams}
\end{table}

Riess et al. (1998) have recognised that weak lensing of high-redshift
Type Ia Supernov\ae~can alter the observed magnitudes, and quote the
findings of Wambsganns et al. (1997) i.e., that the light will be on
average demagnified by 1.7\% at $z=1$ in a universe with a
non-negligible cosmological constant. They state that the size of this
effect is negligible. Perlmutter et al. (1999) assume that the effects
of magnification or demagnification will average out, and that the
most over-dense lines of sight should be rare for their set of 42
high-redshift Supernov\ae. However, they note that the average
(de)amplification bias from integration of the probability
distributions is less than 1\% for redshifts of $z \leq 1$. 

My work has shown that, in the LCDM cosmology, a source at $z=1$ will
have a median deamplification of 3.4\% ($\mu_{\mathrm{peak}}=0.966$),
and 1.3\% at $z=0.5$. Whilst these demagnification factors are still
small, they can result in repositioning of the Supernov\ae~data on the
Hubble diagram with significantly revised values of the cosmological
parameters. In addition, there is a significant probability of
observing highly magnified Supernov\ae; 97\mbox{$\frac{1}{2}$}\% of
all lines of sight display a range of magnifications up to 1.191 at
$z=1$, and this range is equivalent to a dispersion in the magnitudes
of 0.252. The standard deviation in the asymmetrical distributions at
$z = 1$ is 0.078 magnitudes. This is to be compared with the accepted
intrinsic dispersion of 0.17 magnitudes for Type Ia
Supernov\ae~reported by Hamuy et al. (1996) in a set of low-redshift
Supernov\ae. For sources at a redshift of 2, however, Barber et al.'s
(2000) data predict a standard deviation of 0.185 magnitudes from weak
lensing, in excess of the intrinsic dispersion. Clearly, the weak
lensing dispersion will become increasingly important as
Supernov\ae~at greater redshifts are discovered.

In the work by Wambsganss et al. (1997) in the cosmology with
$\Omega_M = 0.4$ and $\Omega_{\Lambda} = 0.6$ they found
$\mu_{\mathrm{low}}=0.951$ and $\mu_{\mathrm{high}}=1.101$ for source
redshifts of $z = 1$, and $\mu_{\mathrm{low}}=0.978$ and
$\mu_{\mathrm{high}}=1.034$ for $z=0.5$. Their median magnification
values were 0.983 for $z = 1$ and 0.993 for $z = 0.5$. These may be
compared directly with the values shown in Table~\ref{mu4C} for the
magnifications in the $\Omega = 0.3,$ $\Omega_{\Lambda} = 0.7$ cosmology I have
investigated, i.e., $\mu_{\mathrm{low}}=0.944$ and
$\mu_{\mathrm{high}}=1.191$ for source redshifts of $z = 1$, and
$\mu_{\mathrm{low}}=0.981$ and $\mu_{\mathrm{high}}=1.070$ for
$z=0.5$. The median values are 0.966 for $z = 1$ and 0.987 for $z =
0.5$.

\begin{figure}
$$\vbox{
\psfig{figure=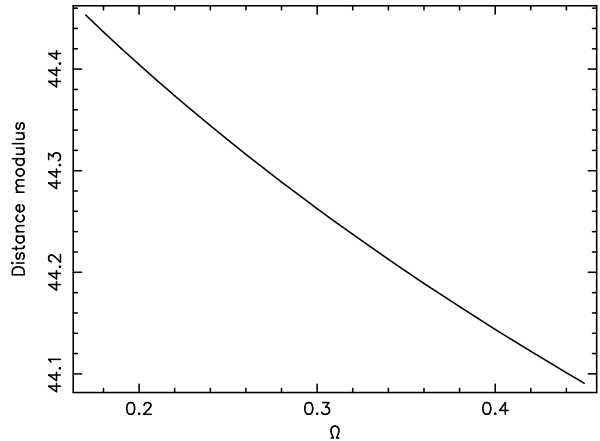,width=8.7truecm,angle=270}
}$$
\caption{The distance modulus for $z=1$ vs. $\Omega_M$, with the 
constraint $\Omega_M +\Omega_{\Lambda} = 1$.}
\label{modulus/Omega}
\end{figure}

Since the most likely effect of weak lensing is a slight
demagnification of the source, it is probable that the originating
data used by Riess et al. (1998) and Perlmutter et al. (1999) may
suffer from such a demagnification, and therefore should be
re-positioned on the Hubble diagram at brighter positions (i.e., lower distance
moduli). This would then point to a slightly different cosmology with a
different deceleration parameter. I plot, in Figure~\ref{modulus/Omega}, the
distance modulus values at $z=1$, for various values of $\Omega_M$,
centred around $\Omega_M = 0.3$, and keeping $\Omega_M +
\Omega_{\Lambda} = 1$. The plot has been generated by solving
equation~\ref{modulus} numerically. 

With no magnification, $\Omega_M = 0.3,$ $\Omega_{\Lambda} = 0.7$,
$h=0.65$, and $z = 1$, the distance modulus is 44.2626. Now assume
that this cosmology is the best fit to the observed data. Then, by
making a redshift 1 Supernova brighter than observed (assuming it was
observed at the peak of the magnification distribution, with
$\mu_{\mathrm{peak}}=0.966$) the distance modulus would be reduced to
44.2250, equivalent to $\Omega_M = 0.33$, and equivalent to changing
$q_0$ from $-0.55$ to $-0.51$. For the rms deviations in the
magnification, the distance modulus changes by $+0.0476/-0.0500$,
equivalent to revised $q_0$ values of $-0.61$ and $-0.49$. For the
$\mu_{\mathrm{high}}$ and $\mu_{\mathrm{low}}$ values at $z=1$, $q_0$
changes to $-0.75$ and $-0.48$.

The conclusion, on the evidence of these data, is that true underlying
cosmologies with $q_0 = -0.51 +0.03/-0.24$ may be interpreted as
having $q_0 = -0.55$, from the use of perfect standard candles
(without intrinsic dispersion), arising purely from the effects of
weak lensing. This adjustment to the value of $q_0$ is almost an order
of magnitude larger than that found by Wambsganss et al. (1997) based
on a cosmology with $\Omega_M = 0.4,$ $\Omega_{\Lambda} = 0.6$, and
the dispersion in the values is slightly larger than their
findings. 

\section{SUMMARY AND CONCLUSIONS}

I have summarised the results for the magnification distributions for
the different cosmologies for different source redshifts as obtained
by Barber et al. (2000) in Table~\ref{mu4C}. At high redshift, the
LCDM cosmology produces the highest magnifications, the broadest
distribution curves, and the lowest peak values. For sources at
$z=3.6$ in the LCDM cosmology, 97\mbox{$\frac{1}{2}$}\% of all lines
of sight have magnification values up to 1.850. (The maximum
magnifications, not quoted here, depended on the choice of the minimum
softening in the code, although the overall distributions were very
insensitive to the softening.) The rms fluctuations in the
magnification (about the mean) were as much as 0.191 in this
cosmology, for sources at $z=3.6$.  Even for sources at $z=0.5$
there is a measurable range of magnifications in all the cosmologies.

The immediate implication is the likely existence of a bias in
observed magnitudes of distant objects, and a likely dispersion for
standard candles, for example, Type Ia Supernov\ae~at high
redshift. In particular, the weak lensing dispersion from
97\mbox{$\frac{1}{2}$}\% of the lines of sight expected in the peak
magnitudes for Type Ia Supernov\ae~at redshifts of 1 may be as much as
0.252 magnitudes. The standard deviations in the distributions
are 0.078 magnitudes for $z = 1$ and 0.185 magnitudes for $z =
2$. These values are to be compared with the accepted dispersion of
0.17 magnitudes, so that we should expect to see an increasing
dispersion for the peak magnitudes of Supernov\ae~as they are
discovered at higher redshifts.

I have made use of the magnification statistics from the LCDM cosmology, to
reanalyse the results for the cosmological parameters determined from
the high-redshift Type Ia Supernov\ae~data. (Both Riess et al., 1998,
and Perlmuter et al., 1999, point to cosmologies with parameters close
to those of the LCDM model, i.e., $\Omega_M = 0.3$, $\Omega_{\Lambda}
= 0.7$.) 

The dispersions in the magnification are monotonically increasing with
redshift, and have rms fluctuations about the mean of 0.045 for $z =
0.99$, and 0.016 for $z = 0.49$. These have been translated into
variations in distance modulus, which, in turn, suggest a different
cosmological model for the data. On the assumption that $\Omega_M =
0.3$, $\Omega_{\Lambda} = 0.7$ and $q_0 = -0.55$ is the cosmology
which fits the observed data best, my results indicate that the true
underlying cosmology could have $q_0 = -0.51$ based on the peak
magnification values, with a spread of $+0.03/-0.24$ based on the
computed magnification dispersion. This would represent a significant
adjustment to the cosmological parameters.

Wambsganss et al. (1997) found a median demagnification value much
closer to unity (1.7\% below for $z = 1$) in their cosmology with
$\Omega_M = 0.4,$ $\Omega_{\Lambda} = 0.6$ than Barber et al. (2000)
have found (3.4\% below for $z = 1$) in the $\Omega_M = 0.3$,
$\Omega_{\Lambda} = 0.7$ cosmology. Consequently, my adjustment to
$q_0$ is considerably greater. However, the magnification dispersions
give rise to similar dispersions in $q_0$.  Wambsganss et al.'s (1997)
calculated adjustment to $q_0$ is approximately an order of magnitude
smaller than the departure I would suggest.

Fluke, Webster and Mortlock's (2000) estimate for the revised value of
the deceleration parameter ($q_0 = 0.53~+0.96/-0.02$ for the $\Omega_M
= 0.3,$ $\Omega_{\Lambda} = 0.7$ cosmology) in the empty beam
approximation is only half the value of the departure I suggest. The
value of their dispersion can not be easily compared with mine,
because of their use of the empty beam approximation, which gives rise
to a different form for the distribution curve with values of
magnification which are always positive.  Furthermore, their method to
convert the empty beam magnification values (more appropriate to their
method) to filled beam values was not rigorous, so that the resulting
values should be treated with caution.

Both Wambsganss et al. (1997) and Fluke, Webster and Mortlock (2000)
have assumed that the observed Supernov\ae~have magnifications of
unity, so that the median demagnifications when using the filled beam
approach are assumed to reposition the Supernov\ae~at dimmer
magnitudes. Fluke, Webster and Mortlock (2000) have to reposition them
at brighter magnitudes when using the empty beam approximation. I have
adopted the opposite viewpoint. I have assumed that the
Supernov\ae~are observed at their median demagnification values, and
therefore have to be repositioned at brighter magnitudes (i.e.,
reduced distance moduli) to obtain the correct cosmological
parameters. Consequently, the direction of my adjustment to $q_0$ is
opposite to those of the above authors for the filled beam
approximation.

\section*{ACKNOWLEDGMENTS}

I am indebted to the Starlink minor node at the University of Sussex
for the preparation of this paper, and to the University of Sussex for
sponsorship. I have benefited from useful exchanges with Dr. Peter Thomas of the University of Sussex and Dr. Chris Fluke of Swinburne University.

\section*{References}

\baselineskip 0.41cm 
\begin{trivlist} 

\item{Aldering G., Knop R. and Nugent P., 2000, astro-ph, 0001049.}
\item{Barber A. J., Thomas P. A., Couchman H. M. P. and Fluke C. J., 2000, MNRAS, submitted (astro-ph 0002437.}
\item{Carroll, S. M., Press, W. H. and Turner, E. L., 1992, A.R.A.\&A., 30, 499.}
\item{Couchman H. M. P., Barber A. J. and Thomas P. A., 1999, MNRAS, 308, 180.}
\item{Couchman H. M. P., Thomas, P. A., and Pearce F. R., 1995, 
Ap. J., 452, 797.}
\item{Fluke C. J. and Webster R. L., 2000, MNRAS (submitted).}
\item{Fluke C. J., Webster R. L. and Mortlock
D. J., 1999, MNRAS, 306, 567.}
\item{Fluke C. J., Webster R. L. and Mortlock D. J., 2000, MNRAS (submitted).}
\item{Goodwin, S. P., Thomas, P. A., Barber, A. J., Gribbin, J. and Onuora, L. I., 1999, astro-ph 9906187, preprint.}
\item{Hamuy, M., Phillips, M., Maza, J., Suntzeff, N., Schommer, R. and Avil\'{e}s, R., 1995, A.J., 109, 1.}
\item{Hamuy, M., Phillips, M., Maza, J., Suntzeff, N., Schommer, R. and Avil\'{e}s, R., 1996, A.J., 112, 2391.}
\item{Peacock J. A. and Dodds S. J., 1994, MNRAS, 267, 1020.}
\item{Peebles P.J.E., 1993, `Principles of Physical Cosmology,' 
Princeton University Press, ISBN 0-691-07428-3.}
\item{Perlmutter, S., Gabi S., Goldhaber G., Goobar A., Groom D. E., Hook I. M., Kim A. G., Kim M. Y., Lee J. C., Pain R., Pennypacker C. R., Small I. A., Ellis R. S., McMahon R. G., Boyle B. J., Bunclark P. S., Carter D., Irwin M. J., Glazebrook K., Newberg H. J. M., Filippenko A. V., Matheson T., Dopita M. and Couch W. J., 1997, Ap. J.,483, 565.}
\item{Perlmutter S., Aldering G., Goldhaber G., Knop R. A., Nugent P.,
Castro P. G., Deustua S., Fabbro S., Goobar A., Groom D. E., Hook
I. M., Kim A. G., Kim M. Y., Lee J. C., Nunes N. J., Pain R.,
Pennypacker C. R., Quimby R., Lidman C., Ellis R. S., Irwin M.,
McMahon R. G., Ruiz-Lapuente P., Walton N., Schaefer B., Boyle B. J.,
Filippenko A.  V., Matheson T., Fruchter A. S., Panagia N., Newberg H.
J. M., Couch W. J., and Project T. S. C., 1999, Ap. J., 517, 565.}
\item{Phillips, M., 1993, Ap.J., 413, L105.}
\item{Riess A. G., Filippenko A. V., Challis P., Clocchiatti A.,
Diercks A., Garnavich P. M., Gilliland R. L., Hogan C. J., Jha S.,
Kirshner R. P., Leibundgut B., Phillips M. M., Reiss D., Schmidt
B. P., Schommer R. A., Smith R. C., Spyromilio J., Stubbs C., Suntzeff
N. B. and Tonry J., 1998, A. J., 116, 1009.}
\item{Riess, A. G., Filippenko, A. V., Li, W. and Schmidt B. P., 1999, astro-ph 9907038, preprint.}
\item{Riess, A., Press, W. and Kirshner, R., 1995, Ap. J., 438, L17.}
\item{Riess, A., Press, W. and Kirshner, R., 1996, Ap. J., 473, 88.}
\item{Schneider P., Ehlers J., and Falco E. E., 1992, `Gravitational 
Lenses,' Springer-Verlag, ISBN 0-387-97070-3.}
\item{Viana P. T. P., and Liddle A. R., 1996, MNRAS, 281, 323.}
\item{Wambsganss J., Cen R., and Ostriker J., 1998, Ap. J., 494, 29.}
\item{Wambsganss J., Cen R., Xu G. and Ostriker
J. P., 1997, Ap. J., 475, L81.}

\end{trivlist}
\begin{description} 
\item 
\end{description} 


\end{document}